# Content based Zero-Watermarking Algorithm for Authentication of Text Documents

Zunera Jalil[1], Anwar M. Mirza[1] and Maria Sabir[2]

[1]FAST National University of Computer and Emerging Sciences, Islamabad, Pakistan

[2]Air University, Islamabad, Pakistan

*Abstract*- **Copyright protection and authentication of digital contents has become a significant issue in the current digital epoch with efficient communication mediums such as internet. Plain text is the rampantly used medium used over the internet for information exchange and it is very crucial to verify the authenticity of information. There are very limited techniques available for plain text watermarking and authentication. This paper presents a novel zero-watermarking algorithm for authentication of plain text. The algorithm generates a watermark based on the text contents and this watermark can later be extracted using extraction algorithm to prove the authenticity of text document. Experimental results demonstrate the effectiveness of the algorithm against tampering attacks identifying watermark accuracy and distortion rate on 10 different text samples of varying length and attacks.**

*Keywords-watermarking; copyright protection; authentication; security; algorithm*

## I. INTRODUCTION

Copyright protection and authentication of digital contents has gained more importance with the increasing use of internet, e-commerce, and other efficient communication technologies. Besides, making it easier to access information within a very short span of time, it has become difficult to protect copyright of digital contents and to prove the authenticity of the obtained information. Digital contents mostly comprises of text, image, audio, and video. Authentication and copyright protection of digital images, audio, and video has been given due thought by the researchers in past. However, authentication and copyright protection of plain text has been neglected. Most of the digital contents like websites, e-books, articles, news, chats, SMS, are in the form of plain text.

The threats of illegal copying, tampering, forgery, plagiarism, falsification, and other forms of possible sabotages need to be specifically addressed. Digital watermarking is one of the solutions which have been used to authenticate and to protect digital contents. Digital watermarking methods are used to identify the original copyright owner (s) of the contents which can be an image, a plain text, an audio, a video or a combination of all.

A digital watermark can be described as a visible or an invisible, preferably the latter, identification code that permanently is embedded in the data. It means that unlike conventional cryptographic techniques, it remains present within the data even after the decryption process [1].

A text, being the simplest mode of communication and information exchange, brings various challenges when it comes to copyright protection and authentication. Any changes on text should preserve the value, usefulness, meaning, and grammaticality of the text. Short documents are more difficult to protect and authenticate since a simple analysis would easily reveal the watermark, thus making text insecure.

In image, audio, and video watermarking the limitations of Human Visual and/or Human Auditory System are exploited for watermark embedding along with the inherent redundancy. It is difficult to find such limitations and redundancy in plain text, since text is sensitive to any modification required to be made for watermark embedding.

Text is easier to copy, reproduce and tamper as compared with images, audio and video. Text being a specialized medium requires specialized copyright protection and authentication solutions. Traditional watermarking algorithms modify the contents of the digital medium to be protected by embedding a watermark. This traditional watermarking approach is not applicable for plain text. A specialized watermarking approach such as zero-watermarking would do the needful for text. In this paper, we propose a novel zero-watermarking algorithm which utilizes the contents of text itself for its authentication. A zero-watermarking algorithm does not change the characters of original data, but utilize the characters of original data to construct original watermark information [2-3].

The paper is organized as follows: Section 2 provides an overview of the previous work done on text watermarking. The proposed embedding and extraction algorithm are described in detail in section 3. Section 4 presents the experimental results for the tampering (insertion, deletion and re-ordering) attacks with different keywords on. Performance of the proposed algorithm is evaluated by co multiple text samples. The last section concludes the paper along with directions for future work.

## II. PREVIOUS WORK

Text watermarking for authentication of text documents is an important area of research; however, the work done in this domain in past is very inadequate. The work on text watermarking initially started in 1991. A number of text watermarking techniques have been proposed since then. These include text watermarking using text images, synonym based, pre-supposition based, syntactic tree based, noun-verb based, word and sentence based, acronym based, typo error based methods etc.

The previous work on digital text watermarking can be classified in the following categories; an image based approach, a syntactic approach, a semantic approach and the structural approach. Description of each category and the work done accordingly is as follows:

*A. An Image-Based Approach*

In image based approach towards text watermarking, the image of text is takes as source for watermark embedding. Brassil, et al. were the first to propose a few text watermarking methods utilizing text image[4]-[5]. Later Maxemchuk, et al. [6]-[8] analyzed the performance of these methods. Low, et al. [9]-[10] further analyzed the efficiency of these methods. The first method was the line-shift algorithm which moves a line upward or downward (left or right) based on watermark bit values. The word-shift algorithm used the inter-word spaces to embed the watermark. The last method was the feature coding algorithm in which specific text features are tampered to encode watermark bits in the text.

Huang and Yan [11] proposed an algorithm based on an average inter-word distance in each line. The distances are adjusted according to the sine-wave of a specific phase and frequency. The feature and the pixel level algorithms were also developed which mark the documents by modifying the stroke features such as width or serif [12].

Text watermarking algorithms using binary text image are not robust against re-typing attack and have limited applicability. Authentication of text becomes easy with text image, but it is mostly impractical to treat text as an image. In some algorithms, watermark can be destroyed by a simple OCR (Optical Character Recognitions) analysis. The use of OCR obliterate the changes made to the spaces, margins and fonts of a text to embed watermark.

*B. A Syntactic Approach*

In this approach towards text watermarking, the syntactic structure of text is used to embed watermark. Mikhail J. Atallah, et al. first proposed the natural language watermarking scheme by using syntactic structure of text [13]-[14] where the syntactic tree is built and transformations are applied to it in order to embed the watermark keeping all the properties of text intact. The NLP techniques are used to analyze the syntactic and the semantic structure of text while performing any transformations to embed the watermark bits.

Hassan et al. performed morpho-syntactic alterations to the text to watermark it [15]. The text is first transformed into a syntactic tree diagram where text hierarchy and dependencies are analyzed to embed watermark bits. Hassan et al. provided an overview of available syntactic tools for text watermarking [16].

Text watermarking by using syntactic structure of text and natural language processing algorithms, is an efficient approach for text authentication and copyright protection but progress in this domain is slower than the requirement. NLP is an immature area of research so far and using in-efficient algorithms, efficient results in text watermarking cannot be obtained.

*C. A Semantic Approach*

The semantic watermarking schemes focus on using the semantic contents of text to embed the watermark. Atallah et al. were the first to propose the semantic watermarking schemes in the year 2000 [17]-[19]. Later, the synonym substitution method was proposed, in which watermark is embedded by replacing certain words with their synonyms [20]. Xingming, et al. proposed noun-verb based technique for text watermarking [21] where nouns and verbs in a sentence are parsed using grammar parser and semantic networks. Later Mercan, et al. proposed an algorithm of the text watermarking by using typos, acronyms and abbreviation to embed watermark [22]. Algorithms were developed to watermark the text using the linguistic semantic phenomena of presuppositions [23] by observing the discourse structure, meanings and representations. The text pruning and the grafting algorithms were also developed in the past. The algorithm based on text meaning representation (TMR) strings has also been proposed [24].

The text watermarking, based on semantics, is language dependent. The synonym based techniques are not resilient to the random synonym substitution attacks. Sensitive nature of some documents e.g. legal documents, poetry and quotes do not allow us to make semantic transformations randomly because in these forms of text a simple transformation sometimes destroys both the semantic connotation and the value of text[25].

*D. A Structural Approach*

This is the most recent approach used for copyright protection of text documents. A text watermarking algorithm for copyright protection of text using occurrences of double letters (aa-zz) in text to embed the watermark has recently been proposed [25]. The algorithm is a blend of encryption, steganography and watermarking. However, groups are formed by using full stop period in this algorithm. Text like poetry, quotes, web contents, legal document may not essentially contain full stops; which makes this algorithm inapplicable to all types of text. To overcome the shortcomings of this algorithm, another algorithm which use preposition besides double letters to watermark text has been proposed [26].

The structural algorithms are not applicable to all types of text documents and are not designed specifically to solve authentication problem; hence we propose a zero-watermarking algorithm which incorporates the contents of text for its protection.

## III. PROPOSED ALGORITHM

The semantic and syntactic watermarking algorithms developed in past for plain text embed a watermark in the host text document itself which results in text quality, meaning and value degradation. We propose a zero-watermarking approach in which the host text document is not altered to embed watermark, rather the characteristics of text are utilized to generate a watermark. The watermark is fragile in nature and is used to authenticate text documents. The watermark generation and extraction process is illustrated is fig. 1. Watermark is registered with the Certifying Authority (CA) and is used is the extraction algorithm to authenticate text document.

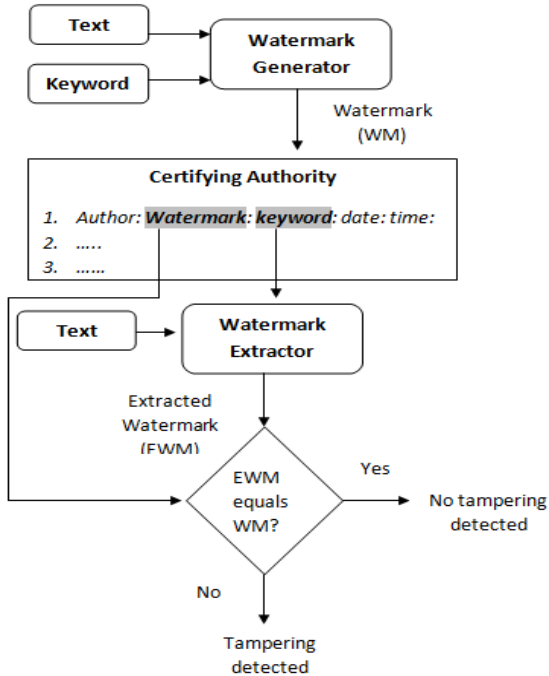

Fig 1: Overview of Watermark Generation and Extraction Processes

The proposed algorithm utilizes the contents of text to protect it. A keyword from the text is selected based on author choice and a watermark is generated based on the length of proceeding and next word length, to and from the keyword occurrences in text. This process is illustrated in fig. 1, where 'is' is the keyword and based on text contents, a watermark is generated.

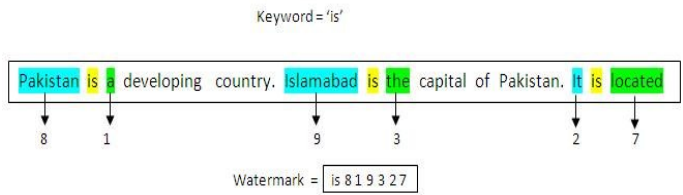

Fig 2: Watermark Generation

It is a zero-watermarking scheme, since watermark is not actually embedded in the text itself; rather it is generated by using the characteristics of text. The watermarking process involves two stages: (1) embedding algorithm and (2) extraction algorithm. Watermark embedding is done by the original author and extraction done later by a Certifying Authority (CA) to prove ownership. A trusted certifying authority is an essential requirement in this algorithm with whom, the original copyright owner registers his/her watermark. Whenever the content/text ownership is in question, this trusted third party acts as a decision authority.

### A. Embedding Algorithm

The algorithm which embeds the watermark in the text is called embedding algorithm. The watermark embedding algorithm requires original text file as input and keyword is selected by the original author/copyright owner. Keyword should be word having frequent occurrence in the text. A watermark is generated as output by this algorithm. This watermark is then registered with the certifying authority along with the original text document, author name, keyword, current date and time. The algorithm proceeds as follows:

```
1. Read T_O.
2. Count Occurrence of each word in T_O.
3. Select KW based on occurrence frequency
4. KWCOUNT = Total occurrence count of KW in
   text T_O
5. for i=1 to KWCOUNT, repeat step 6 to 8.
6.     WM [j] = length (P_i)
7.     WM [j+1] = length (N_i)
8.     i=i+1 and j=j+1
9. Output WM

T_O= Original text; KW=keyword; KWCOUNT= keyword count ;
WM= Watermark; P_i= 'Proceeding word' of the ith occurrence of
keyword (KW); N_i= 'Next word' of the ith occurrence of keyword
(KW)
```

The original text ($T_O$) is first obtained from the author and occurrence frequency of each word in text is analyzed. A keyword is selected by the author which is typical a word with maximum occurrence count in text. The proceeding and next word length for all occurrence of keyword in text is analyzed and a numeric watermark is generated. This watermark is then registered with the CA with current date and time.

### B. Extraction Algorithm

The algorithm which extracts the watermark from the text is called extraction algorithm. The proposed extraction algorithm takes the plain text and keyword as input. The text may be attacked or un-attacked. The watermark is generated from the text by the extraction algorithm and is then, compared with the original watermark registered with the CA. We have also recorded author name, current date and time with the CA. Multiple watermark registration conflicts with CA can be resolved by keeping record of time and date. The author having former registration entry will be regarded as the original author.

The watermark will be accurately detected by this algorithm in the absence of attack on text, and text document will be called authentic text without tampering. The watermark will get distorted in the presence of tampering attacks with text. Tampering can be insertion, deletion, paraphrasing or re-ordering of words and sentences in text. The extraction algorithm is as follows:

```
1. Read T_O or T_A, WM and KW.
2. Count frequency of KW in given text.
3. KWCOUNT = Total occurrence count of KW in
   text
4. for i=1 to KWCOUNT, repeat step 5 to 7.
5.    EWM [ j ] = length (P_i)
6.    EWM [j+1] = length (N_i)
7.    i=i+1 and j=j+1
8. if (EWM not equals WM)
      Tamper = YES
9. Output EWM.

T_O= Original text;  T_A= Attacked text; KW=keyword;
KWCOUNT= keyword count ;  EWM= Extracted Watermark;
P_i= 'Proceeding word' of the ith occurrence of keyword (KW);
N_i= 'Next word' of the ith occurrence of keyword (KW)
```

## IV. EXPERIMENTAL RESULTS

We used 10 samples of variable size text from the data set designed in [25] for our experiments. These samples have been collected from Reuters' corpus, e-books, and web pages. Insertion and deletion of words and sentences was performed at multiple randomly selected locations in text. Table I show the sample label number as in dataset [25], number of words in original text, the insertion and deletion volume, and the number of words in the text after attack.

TABLE I
ORIGINAL AND ATTACKED TEXT SAMPLES WITH INSERTION AND DELETION RATIOS

| Sample No. | Original Text | Attack | | Attacked Text |
|---|---|---|---|---|
| | WC | Insertion | Deletion | Word Count |
| **1 : [SST2]** | 421 | 26% | 25% | 425 |
| **2 : [SST4]** | 179 | 44% | 54% | 161 |
| **3: [MST2]** | 559 | 49% | 25% | 696 |
| **4: [MST4]** | 2018 | 14% | 12% | 2048 |
| **5: [MST5]** | 469 | 57% | 53% | 491 |
| **6: [LST1]** | 7993 | 9% | 6% | 8259 |
| **7: [LST3]** | 1824 | 26% | 16% | 2008 |
| **8: [LST5]** | 16076 | 9% | 5% | 16727 |
| **9: [VLST3]** | 51800 | 11% | 7% | 53603 |
| **10: [VLST5]** | 67214 | 7% | 5% | 68853 |

The number of occurrences of three different keywords "and", "of", and "in" was analyzed in the original and attacked text samples. These keywords were selected because of frequent occurrences in all text samples. Watermark Accuracy Rate (WAR) and Watermark Distortion Rate (WDR) are calculated as per the following formulas:

$$WAR = \frac{\text{Number of characters correctly detected}}{\text{Number of watermark characters}}$$

$$WDR = 1 - WAR$$

The values of WAR ranges between 0 (the lowest) and 1 (the highest) with desirable value close to 1. The values of WDR also ranges between 0 (the highest) and 1 (the lowest) with value close to 0 as desirable value. WAR of the extracted watermark was compared with the original watermark and tamper detection state was analyzed. Table II, III, and IV shows the WAR with keywords 'and', 'of', and 'in' respectively. $WC_O$ and $WC_A$ indicate the keyword count in original and attacked text respectively.

TABLE II
ACCURACY OF EXTRACTED WATERMARK WITH KEYWORD 'AND'

| Sample No. | 'and' | | Tamper Detection | WAR |
|---|---|---|---|---|
| | $WC_O$ | $WC_A$ | | |
| 1 : [SST2] | 12 | 10 | Yes | 0.1538 |
| 2 : [SST4] | 8 | 6 | Yes | 0.4736 |
| 3: [MST2] | 8 | 7 | Yes | 0.2941 |
| 4: [MST4] | 59 | 55 | Yes | 0.1935 |
| 5: [MST5] | 19 | 13 | Yes | 0.3333 |
| 6: [LST1] | 257 | 264 | Yes | 0.1248 |
| 7: [LST3] | 45 | 51 | Yes | 0.4190 |
| 8: [LST5] | 286 | 299 | Yes | 0.1868 |
| 9: [VLST3] | 858 | 915 | Yes | 0.1717 |
| 10: [VLST5] | 1031 | 1053 | Yes | 0.1326 |

It can be observed in table II and III that tampering with text is always detected and low accuracy of watermark indicates that the extent to which text has been attacked. In table IV, the accuracy rate of watermark in sample 4 is 0.2254, even with same frequency counter of keyword 'in' in both original and attacked texts. It depicts the fact that even if the frequency counters of keyword remain intact, the probability of getting same proceeding and next word length for all occurrences of keyword is very low.

TABLE III
ACCURACY OF EXTRACTED WATERMARK WITH KEYWORD 'OF'

| Sample No. | 'of' | | Tamper Detection | WAR |
|---|---|---|---|---|
| | $WC_O$ | $WC_A$ | | |
| 1 : [SST2] | 18 | 23 | Yes | 0.2380 |
| 2 : [SST4] | 7 | 5 | Yes | 0.6153 |
| 3: [MST2] | 9 | 12 | Yes | 0.0952 |
| 4: [MST4] | 64 | 76 | Yes | 0.1582 |
| 5: [MST5] | 7 | 9 | Yes | 0.2352 |
| 6: [LST1] | 237 | 255 | Yes | 0.1526 |
| 7: [LST3] | 38 | 51 | Yes | 0.1125 |
| 8: [LST5] | 571 | 599 | Yes | 0.1680 |
| 9: [VLST3] | 2110 | 2229 | Yes | 0.1323 |
| 10: [VLST5] | 2251 | 2351 | Yes | 0.1407 |

Figure 3 (a), (b), and (c) shows the watermark distortion rate (WDR) with keyword 'and', 'of', and 'in' on all text samples.

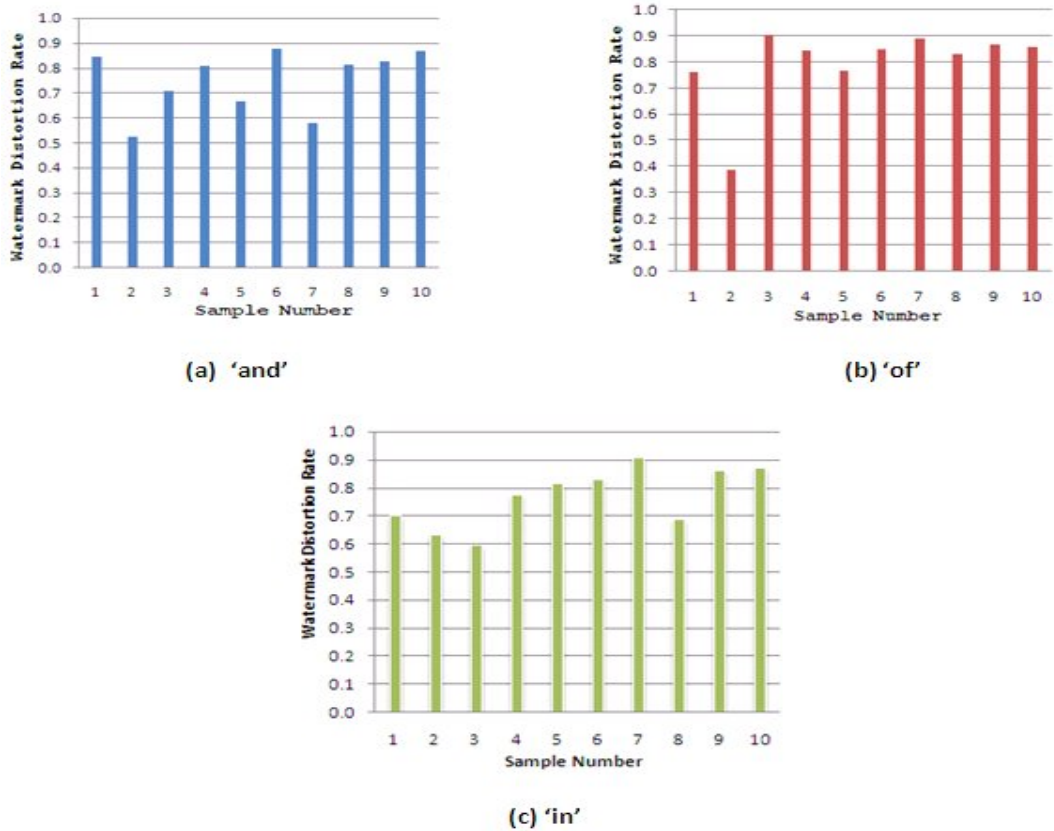

Fig. 3. Watermark distortion rate with keyword (a)'and',(b) 'of', and (c) 'in', with all text samples.

It can be clearly observed that watermark distortion rate is very high even when insertion and deletion volume is low (as in case of sample 8, 9, and 10) for all three keywords. Text is sensitive to any modifications made by the attacker. High distortion rate indicates that the text has been tampered and is not authentic. This proves that the accuracy of watermark gets adversely affected even with minor tampering and watermark fragility proves that text has been attacked.

TABLE IV
ACCURACY OF EXTRACTED WATERMARK WITH KEYWORD 'IN'

| Sample No. | 'in' | | Tamper Detection | WAR |
|---|---|---|---|---|
| | $WC_O$ | $WC_A$ | | |
| 1 : [SST2] | 12 | 11 | Yes | 0.2962 |
| 2 : [SST4] | 4 | 5 | Yes | 0.3636 |
| 3: [MST2] | 11 | 15 | Yes | 0.40 |
| 4: [MST4] | 34 | 34 | Yes | 0.2253 |
| 5: [MST5] | 4 | 14 | Yes | 0.1818 |
| 6: [LST1] | 169 | 162 | Yes | 0.1680 |
| 7: [LST3] | 25 | 27 | Yes | 0.0909 |
| 8: [LST5] | 287 | 291 | Yes | 0.3117 |
| 9: [VLST3] | 904 | 929 | Yes | 0.1354 |
| 10: [VLST5] | 1162 | 1206 | Yes | 0.1266 |

## V. CONCLUSION

The existing text watermarking solutions for text authentication are not applicable under random tampering attacks and on all types of text. With the small volume of attack, it becomes impossible to identify the existence of attack and to prove authenticity of information. We have developed a zero-text watermarking algorithm, which utilizes the contents of text to generate a watermark and this watermark is later extracted to prove the authenticity of text document. We evaluated the performance of the algorithm for random tampering attack in dispersed form on 10 variable size text samples. Results show that our algorithm always detects tampering even when the tampering volume is low.

## ACKNOWLEDGMENT

One of the authors, Ms. Zunera Jalil, 041-101673-Cu-014 would like to acknowledge the Higher Education Commission of Pakistan for providing the funding and resources to complete this work under Indigenous Fellowship Program.

## REFERENCES

[1]. Asifullah Khan, Anwar M. Mirza and Abdul Majid, "Optimizing Perceptual Shaping of a Digital Watermark Using Genetic

Programming", Iranian Journal of Electrical and Computer Engineering, vol. 3, pp. 144-150, 2004.

[2]. Anbo Li, Bing-xian Lin, Ying Chen, "Study on copyright authentication of GIS vector data based on Zero-watermarking", The International Archives of the Photogrammetry, Remote Sensing and Spatial Information Sciences. Vol. VII. Part B4, pp.1783-1786, 2008.

[3]. Zhou, Xinmin , Zhao, Weidong, Wang, Zhicheng, Pan, Li , "Security theory and attack analysis for text watermarking", 2009 International Conference on E-Business and Information System Security, EBISS 2009.

[4]. J. T. Brassil, S. Low, N. F. Maxemchuk, and L. O'Gorman, "Electronic Marking and Identification Techniques to Discourage Document Copying," IEEE Journal on Selected Areas in Communications, vol. 13, no. 8, October 1995, pp. 1495-1504.

[5]. J. T. Brassil, S. Low, and N. F. Maxemchuk, "Copyright Protection for the Electronic Distribution of Text Documents," Proceedings of the IEEE, vol. 87, no. 7, July 1999, pp.1181-1196.

[6]. N. F. Maxemchuk, S. H. Low, "Performance Comparison of Two Text Marking Methods," IEEE Journal of Selected Areas in Communications (JSAC), May 1998. vol. 16 no. 4 1998. pp. 561-572.

[7]. N. F. Maxemchuk, "Electronic Document Distribution," AT&T Technical Journal, September 1994, pp. 73-80. 6.

[8]. N. F. Maxemchuk and S. Low, "Marking Text Documents," Proceedings of the IEEE International Conference on Image Processing, Washington, DC, Oct. 26-29, 1997, pp. 13-16.

[9]. S. H. Low, N. F. Maxemchuk, and A. M. Lapone, "Document Identification for Copyright Protection Using Centroid Detection," IEEE Transactions on Communications, Mar. 1998, vol. 46, no.3, pp 372-381.

[10]. S. H. Low and N. F. Maxemchuk, "Capacity of Text Marking Channel," IEEE Signal Processing Letters, vol. 7, no. 12 , Dec. 2000, pp. 345 -347.

[11]. D. Huang and H. Yan, "Interword distance changes represented by sine waves for watermarking text images," IEEE Trans. Circuits and Systems for Video Technology, Vol.11, No.12, pp.1237-1245, Dec 2001.

[12]. T. Amano and D. Misaki, "A feature calibration method for watermarking of document images," Proceedings of ICDAR, pp.91-94, 1999.

[13]. M. J. Atallah, C. McDonough, S. Nirenburg, and V. Raskin, "Natural Language Processing for Information Assurance and Security: An Overview and Implementations", Proceedings 9th ACM/SIGSAC New Security Paradigms Workshop, September, 2000, Cork, Ireland, pp. 51–65.

[14]. M. J. Atallah, V. Raskin, M. C. Crogan, C. F. Hempelmann, F. Kerschbaum, D. Mohamed, and S.Naik, "Natural language watermarking: Design,analysis, and a proof-of-concept implementation", Proceedings of the Fourth Information HidingWorkshop, vol. LNCS 2137, 25-27 April 2001, Pittsburgh, PA.

[15]. Hassan M. Meral et al., "Natural language watermarking via morphosyntactic alterations", Computer Speech and Language, 23, 107-125, 2009.

[16]. Hasan M. Meral, Emre Sevinç, Ersin Ünkar, Bülent Sankur, A. Sumru Özsoy, Tunga Güngör, Syntactic tools for text watermarking, 19th SPIE Electronic Imaging Conf. 6505: Security, Steganography, and Watermarking of Multimedia Contents, Jan. 2007, San Jose.

[17]. M. Atallah, C. McDonough, S. Nirenburg, and V. Raskin, "Natural Language Processing for Information Assurance and Security: An Overview and Implementations," Proceedings 9th ACM/SIGSAC New Security Paradigms Workshop, September, 2000, Cork, Ireland, pp. 51–65.

[18]. M. Atallah, V. Raskin, C. F. Hempelmann, M. Karahan, R. Sion, U. Topkara, and K. E. Triezenberg, "Natural Language Watermarking and Tamperproofing", Fifth Information Hiding Workshop, vol. LNCS, 2578, October, 2002, Noordwijkerhout, The Netherlands, Springer-Verlag.

[19]. M. Topkara, C. M. Taskiran, and E. Delp, "Natural language watermarking," Proceedings of the SPIE International Conference on Security, Steganography, and Watermarking of Multimedia Contents VII, 2005.

[20]. U. Topkara, M. Topkara, M. J. Atallah, "The Hiding Virtues of Ambiguity: Quantifiably Resilient Watermarking of Natural Language Text through Synonym Substitutions". In Proceedings of ACM Multimedia and Security Conference, Geneva, 2006.

[21]. Xingming Sun, Alex Jessey Asiimwe. Noun-Verb Based Technique of Text Watermarking Using Recursive Decent Semantic Net Parsers.

[22]. Topkara, M., Topraka, U., Atallah, M.J., 2007. Information hiding through errors: a confusing approach. In: Delp III, E.J., Wong, P.W. (Eds.), Security, Steganography, and watermarking of Multimedia Contents IX. Proceedings of SPIE-IS&T Electronic Imaging SPIE 6505. pp. 65050V-1–65050V-12.

[23]. B. Macq and O. Vybornova, "A method of text watermarking using presuppositions," in Proceedings of the SPIE International Conference on Security, Steganography, and Watermarking of Multimedia Contents, January 2007.

[24]. Peng Lu et al., "An optimized natural language watermarking algorithm based on TMR", on proceedings of 9[th] International Conference for Young Computer Scientists, 2009.

[25]. Z. Jalil and A.M. Mirza, "A Novel Text Watermarking Algorithm Based on Double Letters", International Journal of Computer Mathematics. (Submitted)

[26]. Z. Jalil and A. M. Mirza, "A Preposition based Algorithm for Copyright Protection of Text Documents", Journal of the Chinese Institute of Engineers. (Submitted)

Lecture Notes in Computer Science (LNCS) 3612: 958-961, Springer Press, August 2005.